\documentclass[aps,pra,twocolumn,superscriptaddress,showpacs,showkeys,amsmath,amssymb]{revtex4}

\usepackage{amsfonts}
\usepackage{mathrsfs}
\usepackage{latexsym}
\usepackage[cp1251]{inputenc}
\usepackage{graphicx}
\usepackage{dcolumn}
\usepackage{bm}
\usepackage{color}
\RequirePackage{ifthen}
\RequirePackage[pdfstartview=FitH]{hyperref}

\begin{document}

    \title{Large-$N$ expansion for condensation and stability of Bose-Bose mixtures\\ at finite temperatures}
    \author{O.~Hryhorchak}
    \author{V.~Pastukhov\footnote{e-mail: volodyapastukhov@gmail.com}}
    \affiliation{Department for Theoretical Physics, Ivan Franko National University of Lviv,\\ 12 Drahomanov Street, Lviv, Ukraine}

    \date{\today}



    \begin{abstract}
    The two-component mixture of Bose particles with short-range pairwise interaction at finite temperatures in three dimensions is considered. Particularly we examine, by means of the large-$N$ expansion technique, the stability of mixed state below the Bose-Einstein transition point and the temperature dependence of the condensate density for symmetric mixture of Bose gases. The presented analysis reveals the importance of finite-temperature excitations of the non-condensed particles in formation of the phase diagram of two-component Bose systems.      
      
    \end{abstract}

    \maketitle

\section{Introduction}
\label{sec1} \setcounter{equation}{0}
There is a well-established conjecture in the condensed matter physics that mean-field-like theories are capable to capture generic properties of the many-body systems, and more sophisticated treatments applied to the specific problem will only correct some details of the phase diagram. In recent years, however, it was clearly demonstrated the relevance of the beyond-mean-field effects in the formation of macroscopic behavior of ultracold gases even when the interaction between particles is week. These truly quantum fluctuations were found \cite{Petrov} to be responsible for the stabilization of two-component three-dimensional Bose mixtures against a collapse through the formation of stable droplets. The size of these spatially non-uniform objects can vary and depends on the experimental conditions \cite{Cabrera,Semeghini} but each droplet remains to be superfluid and populated by particles of both sorts. Very similar self-bounded structures were predicted in two- \cite{Petrov_Astrakharchik} and one-dimensional \cite{Astrakharchik_Malomed} setups, as well as in the restricted geometries \cite{Zin}. A role of the three-body inter-species attraction in the formation of a spherically symmetric quantum ball in the binary Bose-Einstein condensate (BEC) was elucidated in Ref.~\cite{Gautam}. A first-order-like phase transition \cite{Cheiney} from the low-density bright-solitonic state to quantum liquid droplets were experimentally observed and theoretically analyzed in quasi-one-dimensional mixture of $^{39}$K BECs in two Zeeman states. Further analysis of two-component BECs allowed to propose \cite{Jorgensen} the way of how the beyond mean-field corrections can be measured in these systems and revealed very interesting topological modes, namely, the self-trapped vortex rings \cite{Kartashov} and two-dimensional rotating quantum droplets \cite{Li}.

Such an interest to quantum droplets state stimulated further exploration of properties of uniform binary Bose mixtures in three dimensions by means of variational hypernetted-chain Euler–Lagrange method \cite{Staudinger}, Monte Carlo simulations \cite{Cikojevic_18,Cikojevic_19}, diagrammatic approach \cite{Utesov}. The low-dimensional two-component bosons are also extensively studied both analytically \cite{Konietin,Karle} and numerically \cite{Parisi}. The ground state of the uniform Bose-Bose mixture is well-understood \cite{Larsen,Timmermans} at least in the weakly-interacting regime. It is a phase where two distinct BECs coexist and the spectrum of the elementary excitations is represented by two (phonon-like in the long-wavelength limit) branches \cite{Balabanyan,Oles,Vakarchuk,Rovenchak}. For dimensionalities higher than two, these Bose condensates are robust to the thermodynamic fluctuations smoothly decreasing in temperature and vanish alternately in two distinct critical points. The same concerns the spectrum of collective modes at finite temperatures \cite{Zhang,Armaitis}. On the mean-field level the zero-temperature phase-separation stability condition of a mixed state is provided when $g_{AA}g_{BB}>g^2_{AB}$ (where $g_{ab}$ characterize inter- and intraspecies two-body couplings) and typically inclusion of quantum fluctuations for a model with short-ranged repulsive potentials leads to additional stabilization of the homogeneous system. The finite-temperature predictions based on the mean-field-like \cite{Shi,Van_Schaeybroeck} approximations, or extensions of the Bogoliubov \cite{Colson,Boudjemaa_18,Ota} (see also below) theory for the phase diagram of the Bose-Bose system indicate the stratification of mixture into separate components before the BEC transitions occur. These approaches, however, leave values of the BEC transition temperatures unchanged and totally neglect the impact of the density fluctuations of non-condensed particles. In a case of one-component bosons, account of these enormously developed fluctuations incorporated by the large-$N$ expansion, leads to the non-trivial critical behavior \cite{Hryhorchak_19(2)} of the system and to qualitatively correct shift of the BEC transition temperature \cite{Hryhorchak}. In a context of Bose mixtures, this non-perturbative technique was previously used in Ref.~\cite{Chien} to examine the normal state of the system. The objective of present study is to explore, by means of $1/N$-expansion approach, the interplay of quantum and thermal fluctuations in the two-component system of bosons, in particular, their impact on the stability and thermodynamics of mixture in the superfluid phase.

\section{Model and method}
In what follows we consider two-component (let us say $A$ and $B$) system of mutually interacting Bose gases with point-like inter- and intraspecies two-body potentials. Keeping in mind the large-$N$ expansion method we have to generate $N$ copies of each constituent and to rescale the coupling constants in order to preserve the thermodynamic limit well-defined. More specifically the Euclidean action of our model reads
\begin{eqnarray}\label{S}
	S=\int dx\, \psi^*_{a,\sigma}(x)\left\{\partial_{\tau}-\xi_a\right\}
	\psi_{a,\sigma}(x)\nonumber\\
	-\frac{1}{2N}\int dx\, g_{ab} |\psi_{a,\sigma}(x)|^2|\psi_{b,\sigma'}(x)|^2,
\end{eqnarray}
where summations over the repeating sort $a,b=A,B$ and `flavor' $\sigma,\sigma'=1,\ldots,N$ indices are assumed; $x\equiv(\tau, {\bf r})$ denotes the $3+1$ `position' in `volume' $V/T$ ($T$ is the temperature) with periodic boundary conditions, $\int dx=\int_0^{1/T}d\tau\int_Vd{\bf r}$ and the complex fields $\psi_{a,\sigma}(x)$ describe two-component bosons. The first term in Eq.~(\ref{S}) is the action of two ideal Bose gases with quadratic dispersions $\xi_a=-\hbar^2\nabla^2/2m_a-\mu_a$ and chemical potentials $\mu_a$ that fix the number of particles of each sort, while the second term accounts the zero-range interparticle interactions characterized by the bare couplings $g_{ab}$.

From practical point of view it is more convenient to work with the Hubbard-Stratonovich-transformed action
\begin{eqnarray}\label{S_prime}
	S=\int dx\, \psi^*_{a,\sigma}(x)\left\{\partial_{\tau}-\xi_a-i\varphi_a(x)\right\}
	\psi_{a,\sigma}(x)\nonumber\\
	-\frac{N}{2}\int dx \, g^{-1}_{ab}
	\varphi_{a}(x)\varphi_{b}(x),
\end{eqnarray}
(where $g^{-1}_{ab}$ is the matrix inverse to $g_{ab}$) that enables to explicitly classify (by powers of $N$) terms appearing in perturbative calculations. Following the prescription previously used \cite{Hryhorchak_19} for one-component bosons, we must single out uniform parts of auxiliary real fields $\varphi_a(x)=\varphi_{0a}+\tilde{\varphi}_a(x)$ [where $\int_Vd{\bf r}\tilde{\varphi}_a(x)=0$] and in phases with partially or fully broken global $U(1)\times U(1)$ (for each `flavor' $\sigma$) symmetry to separate the condensate $\psi_{a,\sigma}(x)=\phi_{a,\sigma}+\tilde{\psi}_{a,\sigma}(x)$, where mean value of bosonic fields is $\langle \psi_{a,\sigma}(x)\rangle=\phi_{a,\sigma}$. Then by integrating out fluctuation fields $\tilde{\psi}_{a,\sigma}(x)$ and $\tilde{\varphi}_a(x)$ we are left with the grand thermodynamic potential $\Omega(\mu_a,|\phi_{a,\sigma}|^2, \varphi_{0a})$ which is a function of chemical potentials, uniform parts of the auxiliary fields $\varphi_{0a}$ and condensate densities $|\phi_{a,\sigma}|^2$ of each `flavor' for two sorts of particles. The saddle-point evaluation of integrals over $\phi_{a,\sigma}$ and $\varphi_{0a}$ is equivalent to following identities:
\begin{eqnarray}
\left(\frac{\partial \Omega}{\partial
\varphi_{0a}}\right)_{\mu_a,|\phi_{a,\sigma}|^2}=0, \Rightarrow
i\varphi_{0a}=g_{ab}n_b,
\end{eqnarray}
and from $\left(\frac{\partial \Omega}{\partial
	\phi^*_{a,\sigma}}\right)_{\mu_a,\varphi_{0a}}=0$, we have for arbitrary $a$
\begin{eqnarray}\label{mu_a}
\tilde{\mu}_a\phi_{a,\sigma}-\frac{iT}{V}\int dx \langle \tilde{\psi}_{a,\sigma}(x)\tilde{\varphi}_{a}(x)\rangle=0,
\end{eqnarray}
where we adopted notations $\tilde{\mu}_a=\mu_a-g_{ab}n_b$, $\langle\ldots \rangle$ is the statistical averaging, and denoted, by $Nn_a$, a total density of sort $a$ that, in turn, is fixed by thermodynamic relation $Nn_a=-V^{-1}\left(\partial\Omega/\partial \mu_a\right)$. 

We can now take an advantage of the large-$N$ limit, where the main contributions, which are of order $N$ in this classification, are going from the non-interacting Bose gasses. The path integral over the quadratic in fields $\tilde{\varphi}_a(x)$ part of action $S$, which impacts to terms of order unity in the thermodynamic potential, can be also exactly calculated
\begin{eqnarray}\label{Omega}
\Omega=\frac{VN}{2T}g^{-1}_{ab}\varphi_{0a}\varphi_{0b}-V\tilde{\mu}_a|\phi_{a,\sigma}|^2
\nonumber\\
-TN\sum_{{\bf k},a}\ln\left[1-e^{-\tilde{\xi}_a(k)/T}\right]\nonumber\\
+\frac{T}{2}\sum_{K}\ln \det \left|1+g_{ab}\Pi_{bc}(K)\right|.
\end{eqnarray}
Here, $\tilde{\xi}_a(k)=\hbar^2k^2/2m_a-\tilde{\mu}_a$, the `four-vector' $K=(\omega_k, {\bf k})$ stands for the bosonic Matsubara frequency and wave-vector, and note that in both sums term ${\bf k}=0$ is omitted. The impact of both quantum and finite-temperature fluctuations is contained in the polarization operator $\Pi_{ab}(K)=\delta_{ab}\left[\Pi^{(0)}_{a}(K)+\Pi^{(T)}_{a}(K)\right]$, which is a diagonal matrix with elements given by sum of two terms (no summation over $a$ and $\sigma$)
\begin{eqnarray}\label{Pi_0}
\Pi^{(0)}_{a}(K)=\frac{|\phi_{a,\sigma}|^2}{\tilde{\xi}_a(k)-i\omega_k}+{\rm c.c.},
\end{eqnarray}
is the condensate contribution to the polarization operator and
\begin{eqnarray}\label{Pi_T}
\Pi^{(T)}_{a}(K)=\frac{1}{V}\sum_{{\bf q}}\frac{n(\tilde{\xi}_a(q)/T)-n(\tilde{\xi}_a(|{\bf q}+{\bf k}|)/T)}{\tilde{\xi}_a(|{\bf q}+{\bf k}|)-\tilde{\xi}_a(q)-i\omega_{k}},
\end{eqnarray}
(here, $n(x)=1/(e^x-1)$ stands for the Bose distribution) is a `particle-hole' bubble that represents density fluctuations of the thermally excited from condensate particles. Thus, the thermodynamic potential (\ref{Omega}), which is basic for our large-$N$ analysis of two-component Bose mixtures, incorporates the correct description of the low-temperature region, where it reproduces Bogoliubov's theory, and the simplest inclusion of the finite-temperature fluctuations, which are enormously developed close to the BEC transition point.

The presented formulation is very convenient for the obtaining of various exact relations. Particularly, a singe differentiation of {\it any} vertex function with respect to $\varphi_{0a}$, $\phi_{a,\sigma}$ (or $\phi^*_{a,\sigma}$) increases the number of an appropriate outgoing lines, carrying zero four-momentum, from this vertex by one. This works in such a way that the zero-momentum normal
\begin{eqnarray}\label{Sigma_n}
\Sigma_{\psi^*_{a,\sigma}\psi_{a',\sigma'}}(0)-\tilde{\mu}_{a}\delta_{a,a'}\delta_{\sigma,\sigma'}=\frac{\partial^2}{\partial \phi^*_{a,\sigma} \partial \phi_{a',\sigma'}}\frac{\Omega}{V},
\end{eqnarray}
and anomalous 
\begin{eqnarray}\label{Sigma_an}
\Sigma_{\psi_{a,\sigma}\psi_{a',\sigma'}}(0)=\frac{\partial^2}{\partial \phi_{a,\sigma} \partial \phi_{a',\sigma'}}\frac{\Omega}{V},
\end{eqnarray}
self energies of the single-particle Green's function can be straightforwardly related to density of the thermodynamic potential. Combining these two equations and saddle-point condition $\left({\partial \Omega}/{\partial\phi^*_{a,\sigma}}\right)_{\mu_a,\varphi_{0a}}=0$, we can obtain set of the Hugenholtz-Pines relations for considered Bose mixture (assuming all $\phi_{a,\sigma}$ are real)
\begin{eqnarray}\label{H-P}
\Sigma_{\psi^*_{a,\sigma}\psi_{a,\sigma}}(0)-\tilde{\mu}_{a}=\Sigma_{\psi_{a,\sigma}\psi_{a,\sigma}}(0),
\end{eqnarray}
which are the obvious generalization of the two-component case \cite{Nepomnyashchii}. It is also easy to relate an exact polarization operator $\Pi_{ab}(K)$, which enters 
equation for the inverse matrix of two-point correlation function $\langle \varphi_{a,K} \varphi_{b,-K}\rangle$ ( where $\varphi_{a,K}$ is the four-dimensional Fourier transform of collective field $\tilde{\varphi}_{a}(x)$)
\begin{eqnarray}\label{varphi_varphi}
\langle \varphi_{K} \varphi_{-K}\rangle^{-1}_{ab}=N\left[g^{-1}_{ab}+\Pi_{ab}(K)\right],
\end{eqnarray}
with thermodynamics of the system. In static limit $\omega_k=0$, ${\bf k} \rightarrow 0$, the above equation reads 
\begin{eqnarray}\label{varphi_varphi_stlim}
g^{-1}_{ab}+\Pi_{ab}(0)=\frac{1}{N}\frac{\partial^2}{\partial \varphi_{0a} \partial \varphi_{0b}}\frac{\Omega}{V},
\end{eqnarray}
which allows to rewrite matrix elements of polarization operator in terms of derivatives of densities with respect to the mean-field shifted chemical potentials $\Pi_{ab}(0)=\partial n_a/\partial \tilde{\mu}_b$. The latter identity is consistent with the compressibility sum rule $S_{ab}(0)=\partial n_a/\partial \mu_b$ for the mixture dynamic structure factor.

\section{Results for symmetric mixture}
In the following we focus on case of symmetric mixture, when mass of particles of both sorts, coupling constants of inter-species interaction and densities are equal, i.e., $m_A=m_B=m$, $g_{AA}=g_{BB}=g$ and $n_A=n_B=n$. The only diversity with fully symmetric mixture is that we assume $g_{AB}$ (denoting by $\tilde{g}$) to be different from $g$. Furthermore, below we assume everywhere that $g>\tilde{g}$. All these simplifications provide that the determinant in thermodynamic potential (\ref{Omega}) separates into product of two factors, namely, $\det \left|1+g_{ab}\Pi_{bc}(K)\right|=[1+g_{+}\Pi(K)][1+g_{-}\Pi(K)]$ [here $\Pi(K)$ is given by Eqs.~(\ref{Pi_0}),(\ref{Pi_T})], and in some sense, the whole two-component system may be thought as a mixture of mutually non-interacting Bose gasses with couplings $g_{\pm}=g\pm \tilde{g}$.

\subsection{Stability}
The thermodynamic stability of the system in mixed homogeneous state is provided by conditions $\partial \mu_a/\partial n_a>0$ and 
\begin{eqnarray}\label{gen_stab_c}
\frac{\partial \mu_A}{\partial n_A}\frac{\partial \mu_B}{\partial n_B}-\frac{\partial \mu_A}{\partial n_B}\frac{\partial\mu_B}{\partial n_A}>0,
\end{eqnarray}
which for symmetric case simplifies to $\partial\mu_A/\partial n_A-\partial\mu_A/\partial n_B>0$. The straightforward calculations of chemical potential
\begin{eqnarray}\label{chem_pot}
\mu=ng_{+}+\frac{T}{NV}\sum_{K}\frac{g+g_{+}g_{-}\Pi(K)}{D(K)}
\frac{1}{\varepsilon_k-i\omega_k},
\end{eqnarray}
with the following evaluation of derivatives with respect to densities finally yield for the stability condition
\begin{eqnarray}\label{stab_c}
g_{-}\left\{1-\frac{2T}{NV}\sum_{K}\frac{g_{+}}{D(K)}
\frac{\varepsilon^2_k}{(\omega^2_k+\varepsilon^2_k)^2}\right\}>0,
\end{eqnarray}
where $D(K)=[1+g_{+}\Pi(K)][1+g_{-}\Pi(K)]$, $\varepsilon_k=\hbar^2k^2/2m$ is the free-particle dispersion and recall that Eq.~(\ref{stab_c}) is obtained for the symmetric mixture. In the limit $N=\infty$ the two-component Bose mixture, as it follows from the present study, is always stable when $g_{-}>0$. For finite $N$, however, particularly for the case $N=1$ which actually is of our current interest the phase separation phenomenon at non-zero temperature can occur even if the mean-field stability condition for symmetric binary Bose systems $g>\tilde{g}$ is satisfied. In general, increase of the two-body repulsion due to the impact of beyond mean-field effects leads to the quantum-mechanical stabilization of the mixture, but for small $g$ (and $\tilde{g}$, respectively) the interplay of quantum and thermal fluctuations especially in a region not too far from the BEC transition point drastically changes the phase diagram. In the limit of weak repulsion, the first-order transition temperature (denoting it by $T_s$) can be computed from the zero-frequency term in (\ref{stab_c}). Fixing the interaction strength by $a^3n$, where $a$ is the $s$-wave scattering length for particles of the same sort, i.e. $g=4\pi\hbar^2a/m$, we find that inequality in Eq.~(\ref{stab_c}) is violated at temperature
\begin{eqnarray}\label{tau_s}
\frac{T_s}{T_0}=1-an^{1/3}\tau_s\left(\frac{g_{-}}{g}\right), 
\end{eqnarray}
where $T_0$ is the BEC temperature of the non-interacting system and dimensionless function $\tau_s(g_{-}/g)$ is presented in Fig.~1.
\begin{figure}[h!]
	\centerline{\includegraphics
		[width=0.50\textwidth,clip,angle=-0]{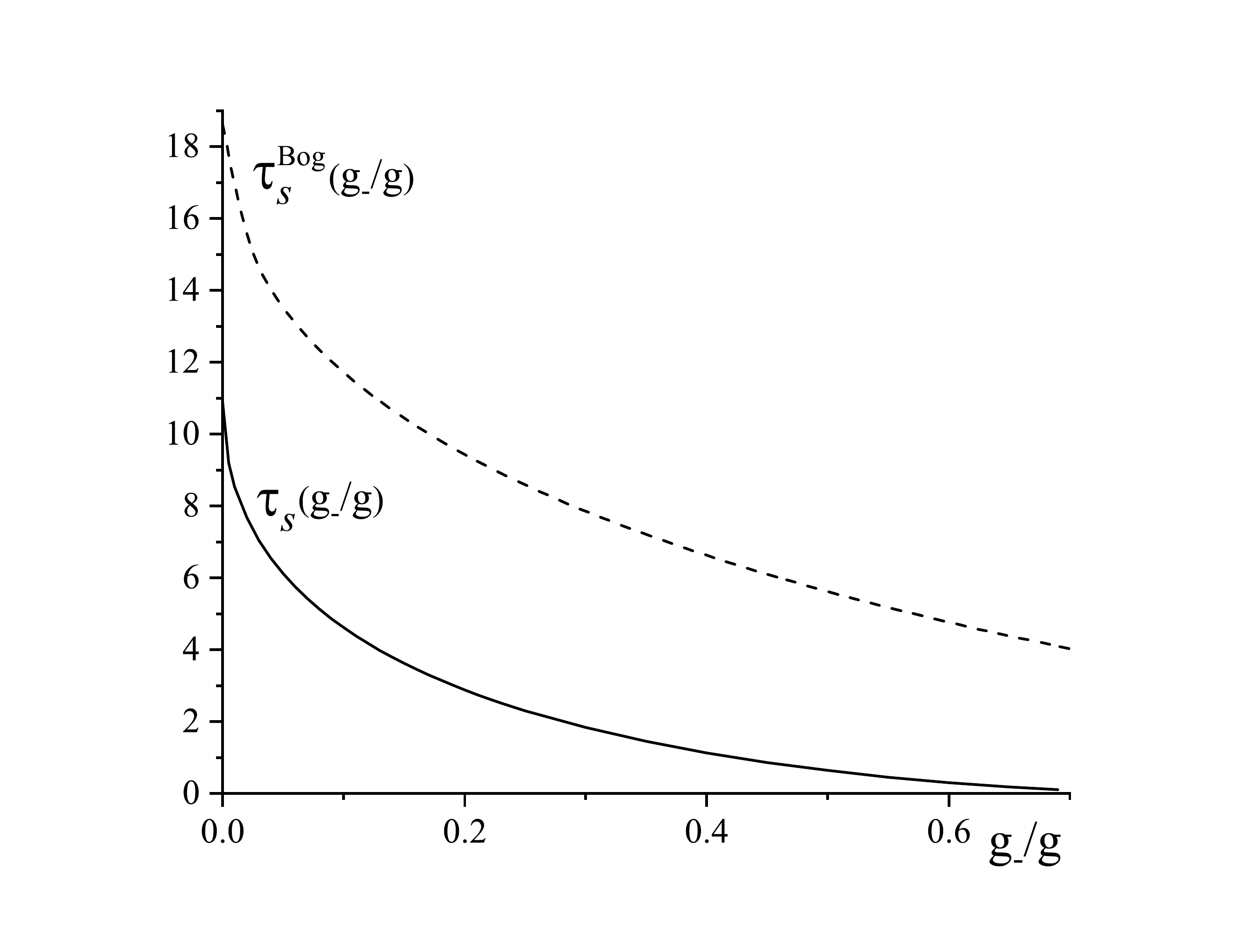}}
	\caption{Dimensionless function that determines the linear at small $an^{1/3}$ $1/N$-shift of the phase-separation temperature $T_s$ [see Eq.~(\ref{tau_s})] (solid line) compared to prediction of Bogoliubov's theory Eq.~(\ref{tau_Bog}) (dashed line).}
\end{figure}
This result suggests that even at very weak interactions between particles $a^3n\sim 10^{-6}\div 10^{-4}$ the characteristic temperature shift $(T_0-T_s)/T_0\sim 10^{-2}\div 10^{-1}$ should be noticeable. In practice, however, this estimation is valid for very small couplings and the phase-separation region, due to full numerical calculations of stability condition (\ref{stab_c}), initially increases with the increase of $g$, reaches maximum at $a^3n=9 \times 10^{-5} $ and then starts to decrease. 
\begin{figure}[h!!]
	\centerline{\includegraphics
		[width=0.50\textwidth,clip,angle=-0]{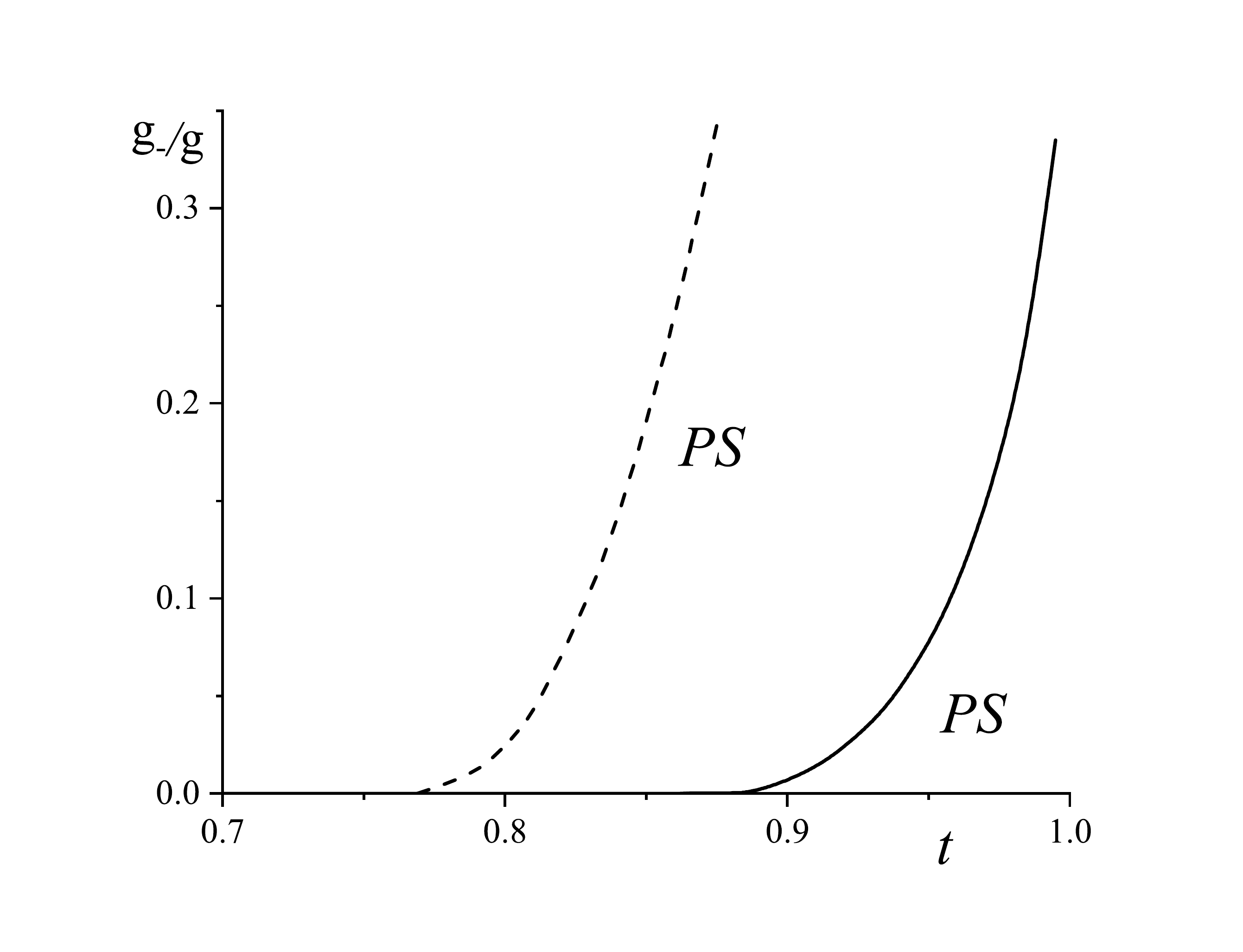}}
	\caption{Example of the phase diagram in dimensionless variables $(g_{-}/g,t=T/T_c)$ for the symmetric Bose-Bose mixture at $a^3n=10^{-4}$. Solid and dashed  lines display our large-$N$ calculations and the result obtained in the Bogoliubov approximation, respectively. The phase separation PS regions are on the right hand sides of the appropriate curves. It is seen that the phase separation could potentially occur at nearly $10\%$ lower temperatures than BEC transition.}
\end{figure}
This behavior is an obvious evidence of the interplay between quantum and thermal fluctuations. Indeed, at very small $g$s the highly developed finite-temperature density fluctuations of the non-condensed particles prevail in the region close to the BEC transition temperature, while the role of the truly quantum beyond mean-field effects becomes to be decisive only when the two-body repulsion is large enough. In Fig.~(2) we have demonstrated typical phase diagram of the two-component symmetric Bose mixture in dimensionless parameters $(g_{-}/g,t)$. Here $t=T/T_c$ is measured in units of the BEC transition temperature, which within the leading-order $1/N$-approximation can be replaced by $T_0$. The value of interaction parameter $a^3n=10^{-4}$ is deliberately chosen to reveal the maximally possible phase-separation region. It is also instructive to compare the large-$N$ phase diagram of this simplified model with that given by the conventional Bogoliubov theory. For these purposes, in the phase-separation (\ref*{stab_c}) condition, an impact of the finite-temperature density fluctuations of the non-condensed particles (\ref{Pi_T}) has to be neglected. The main difference of Bogoliubov's limit is that the two-component system driving from low-temperature condensate phase {\it always} undergoes stratification phenomenon before the BEC transition occurs. In the limit of weak interparticle interaction the leading-order shift of the phase-separation temperature (\ref{tau_s}) is also linear in $an^{1/3}$ with the dimensionless function $\tau_s(g_{-}/g)$ that can be analytically computed
\begin{eqnarray}\label{tau_Bog}
\tau^{\textrm{Bog}}_s\left(\frac{g_{-}}{g}\right)=\frac{16\pi}{3[\zeta(3/2)]^{4/3}}\frac{\left(2-\frac{g_{-}}{g}\right)^2}{1+\sqrt{\frac{g_{-}}{g}}\sqrt{2-\frac{g_{-}}{g}}},
\end{eqnarray}
where $\zeta(z)=\sum_{n\ge 1}1/n^z$ is Riemann zeta function.

\subsection{Bose-Einstein condensation}
The previous section outlined the region of stability for the mixed state of a two-component Bose system at finite temperatures. So we are ready to explore the BEC transition of the symmetric mixture for set of parameters, where it remains stable. Because, as we will find out below, the interaction-induced shift of the BEC critical temperature (at least for $a^3n<4\div 6\times 10^{-2}$) is {\it positive}, while the phase separation always happens {\it below} $T_c$. Therefore not for all values of $g$ and $\tilde{g}$ the two-component system will attain the BEC transition as a homogeneous mixture.

As usual, the transition point is determined by the temperature, where the Bose condensate disappears. Therefore, in order to calculate the critical temperature we must obtain the condensate density (here $n_0=|\phi_{a, \sigma}|^2$ with $a$ and $\sigma$ kept fixed) 
\begin{align}\label{n_0}
	& n_0=n-\frac{1}{V}\sum_{{\bf k}}n\left([\varepsilon_k-\tilde{\mu}]/T\right) \nonumber\\
	& +\frac{T}{2NV}\sum_{K}\frac{g+g_{+}g_{-}\Pi(K)}{D(K)}
	\frac{\partial \Pi(K)}{\partial \tilde{\mu}},
\end{align}
which is valid for symmetric mixture. Working in the adopted approximation we additionally have to take into account the explicit dependence of shifted chemical potential $\tilde{\mu}=\mu-ng_{+}$ on parameter $1/N$ [see Eq.~(\ref{chem_pot})]. It particularly means that we have to expand the second term in Eq.~(\ref{n_0}) up to order $1/N$ and put $\tilde{\mu}=0$ in the third one (after the evaluation of derivative, of course). It is well-known that the $1/N$ expansion predicts \cite{Baym} a correct dependence of the critical-temperature shift on the coupling parameter for Bose gas with weak interparticle repulsion. Moreover, recently we have shown \cite{Hryhorchak} that it also provides a qualitatively correct description of the BEC transition for all couplings predicting a non-monotonous dependence of $T_c$ on gas parameter $a^3n$. Note that the Bogoliubov approach does not provide the shift of the BEC transition temperature in comparison to ideal Bose gas. In general, the critical temperature for the symmetric mixture in the adopted approximation reads
\begin{eqnarray}\label{T_c}
\frac{T_c}{T_0}=1+\frac{1}{N}\tau_c\left(a^3n,\frac{g_{-}}{g}\right), 
\end{eqnarray}
where dimensionless function $\tau_c\left(a^3n,\frac{g_{-}}{g}\right)$ contains all information about intra- and interspecies interactions. Likewise the one-component case, the shift of $T_c$ for considered system is linear in $an^{1/3}$ and independent on ratio $g_{-}/g$ for weakly-repulsive two-body potentials. The increase of interaction strength, as it depicted in Fig.~(3),
\begin{figure}[h!]
	\centerline{\includegraphics
		[width=0.50\textwidth,clip,angle=-0]{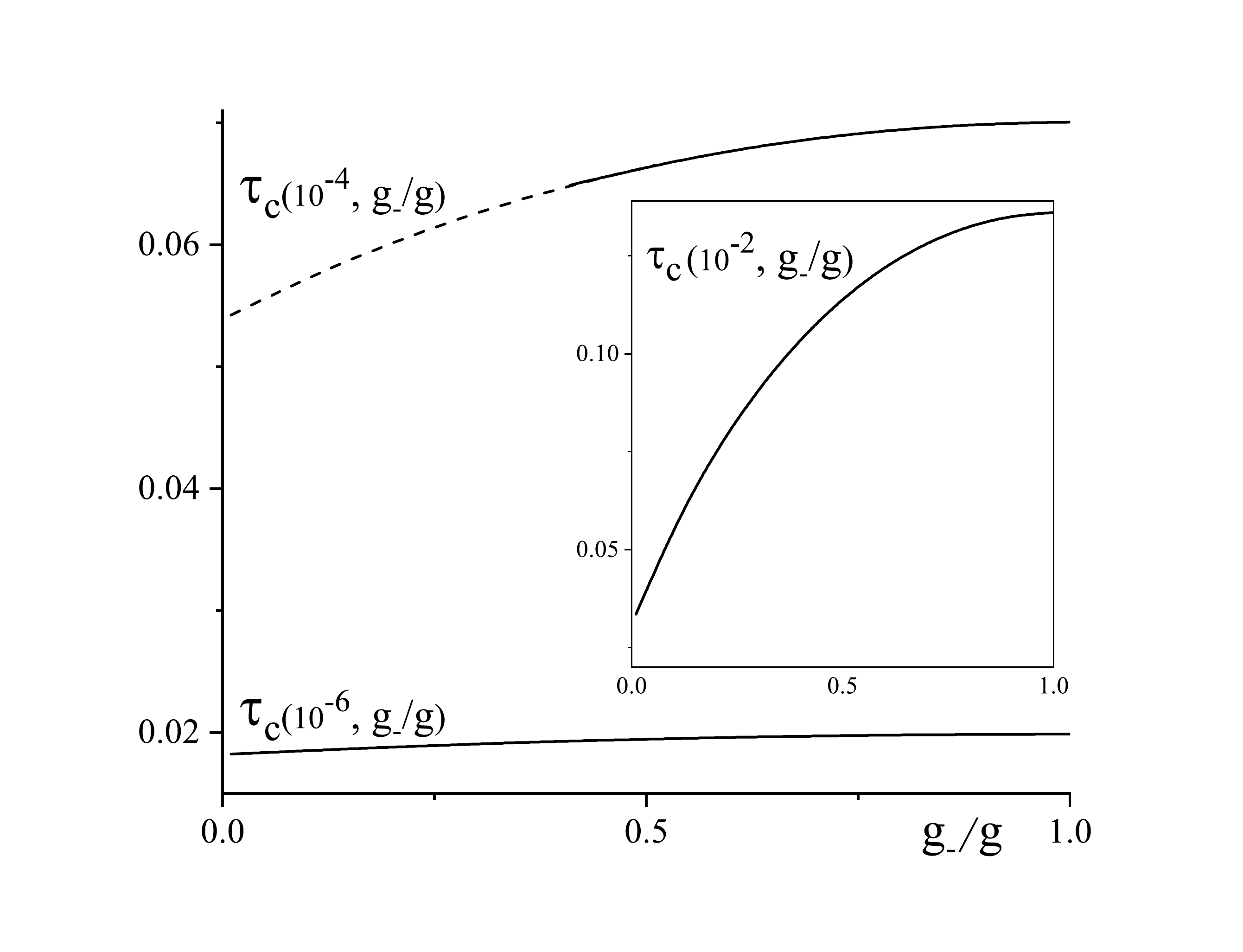}}
	\caption{Dimensionless function $\tau_c\left(a^3n,\frac{g_{-}}{g}\right)$ [see Eq.~(\ref{T_c})] that determines the leading-order shift of the BEC transition temperature at $a^3n=10^{-6}$ (lower line), $a^3n=10^{-4}$ (upper line) and $a^3n=10^{-2}$ (inset). Dashed part of line indicates region, where the mixture is phase-separated [see Fig.~(2)].}
\end{figure}
leads to strengthening of the dependence of function $\tau_c\left(a^3n,\frac{g_{-}}{g}\right)$ on the second argument. Furthermore in Fig.~(3) the region of mechanical instability of the symmetric Bose-Bose mixture at $a^3n=10^{-4}$ is shown (for other curves it is a very narrow region near the origin).

We have also studied the temperature behavior of the Bose condensate for our symmetrical model. Recently we have demonstrated \cite{Hryhorchak_19(2)} the relevance of our large-$N$ treatment by comparison of the calculated Bose condensate and superfluid densities with results of Monte Carlo simulations \cite{Pilati}. Furthermore, the adopted approach also provides \cite{Hryhorchak_19} the qualitative correct behavior of the thermodynamic function in the narrow region of the BEC critical point. Therefore there is a belief that the $1/N$ expansion method will be helpful in the two-component case. For our numerical computations we have chosen two values of gas parameters, $a^3n=10^{-4}$ and $a^3n=10^{-2}$ [see Figs.~(4), (5)], at various ratios $g_{-}/g=0.05, 0.25, 1$ of coupling parameters. 
\begin{figure}[h!]
	\centerline{\includegraphics
		[width=0.50\textwidth,clip,angle=-0]{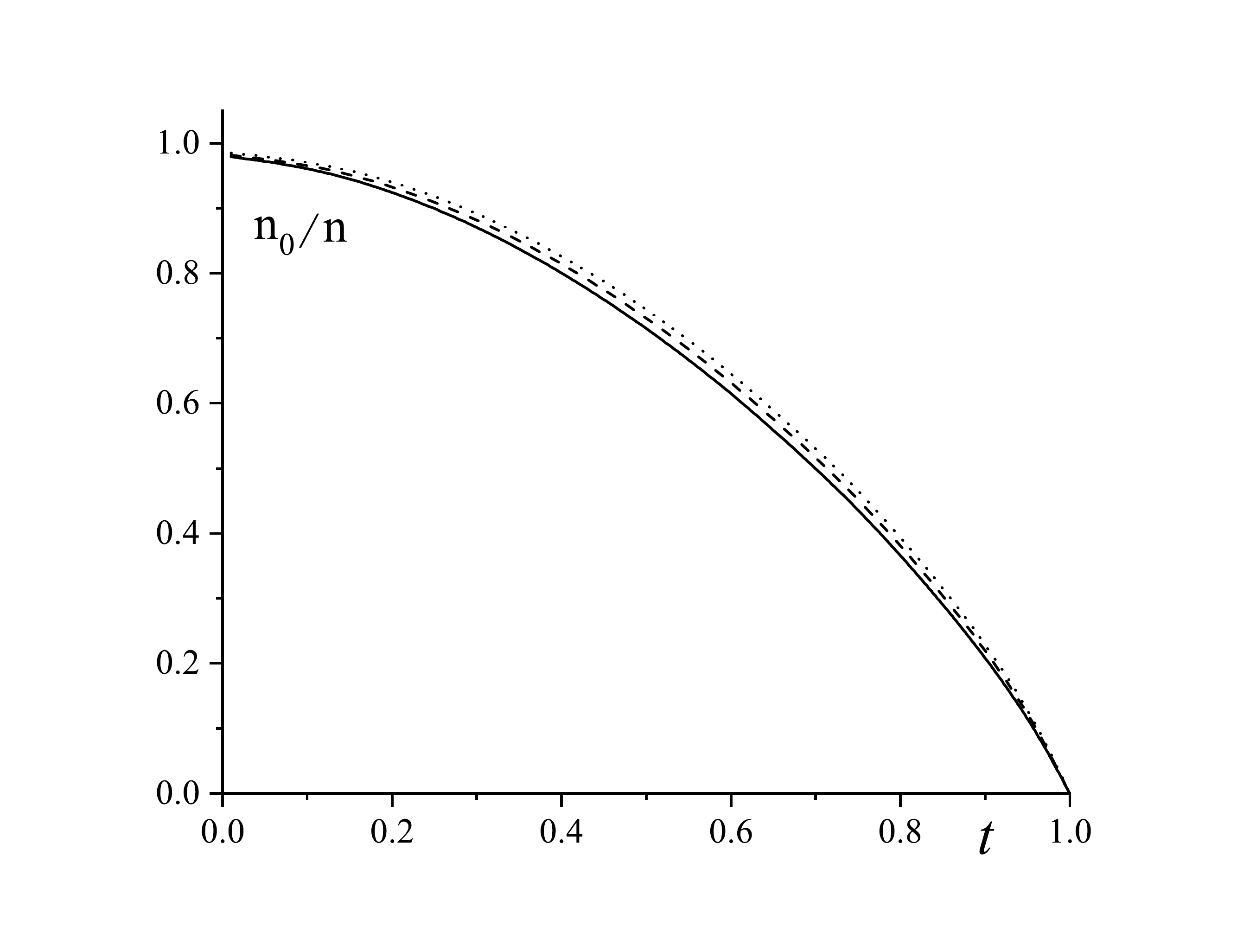}}
	\caption{Temperature dependence of condensate fraction for symmetric mixture at $a^3n=10^{-4}$ for three ratios $g_{-}/g=$$ 0.05$ (solid line), $g_{-}/g=0.25$ (dashed line), $g_{-}/g=1$ (dotted line).}
\end{figure}
\begin{figure}[h!]
	\centerline{\includegraphics
		[width=0.50\textwidth,clip,angle=-0]{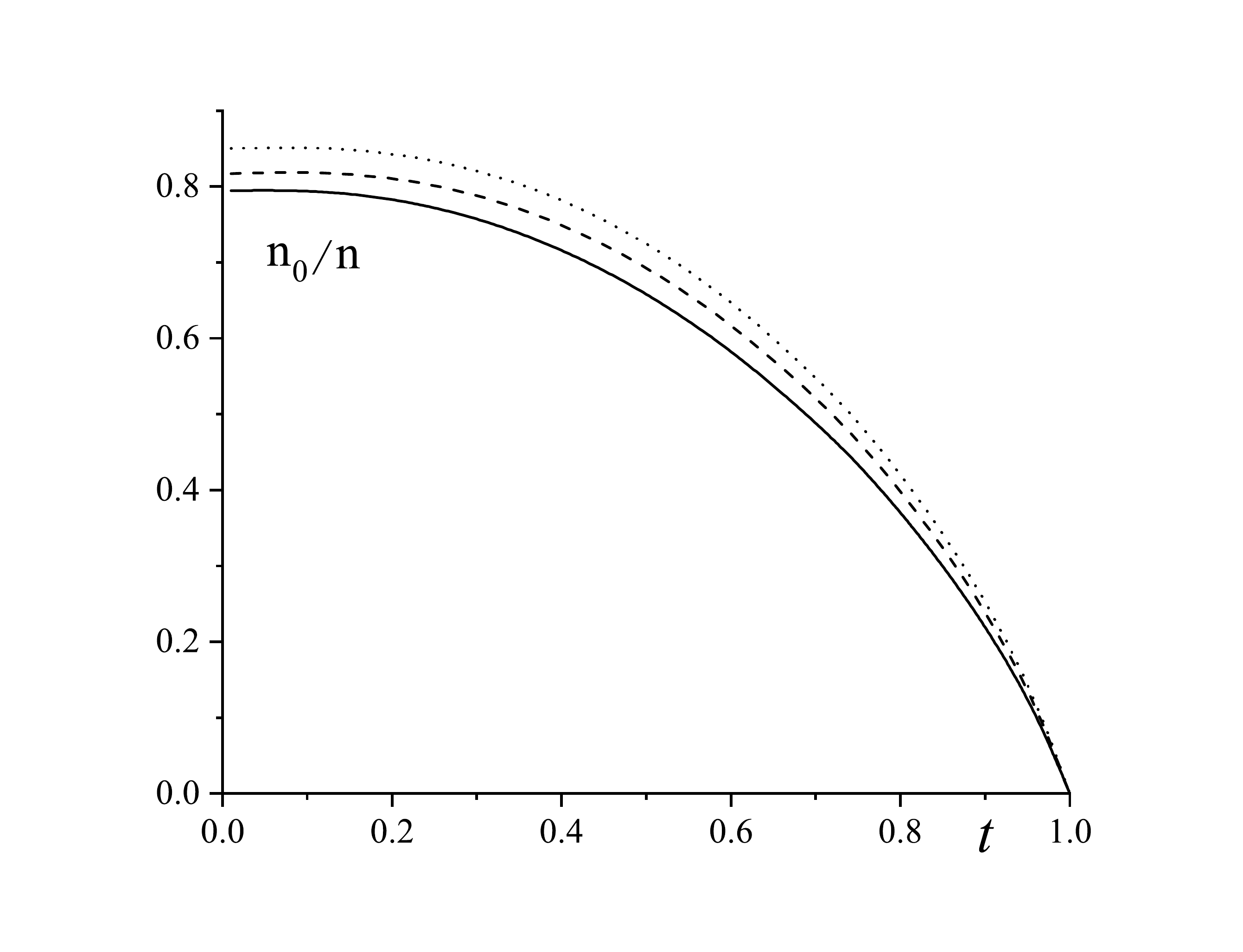}}
	\caption{Condensate fraction at $a^3n=10^{-2}$ (designations of curves are similar to those in Fig.~4).}
\end{figure}
As expected, presence of the repulsive interspecies interaction generally depletes the Bose condensate.

\section{Summary}
In conclusion, we have adapted the large-$N$ expansion technique for the description of two-component Bose systems in the condensate phase. Particularly, applying this method up to the first-order approximation in expansion parameter for a symmetric Bose-Bose mixture, we have demonstrated that the inclusion of density fluctuations of non-condensed particles substantially change the finite-temperature phase diagram of the system providing that the stratification of mixture, predicted by the mean-field and Bogoliubov's theories, is mostly suppressed. Considering system of two-component equal-mass bosons of identical densities and interspecies repulsive interaction, we have numerically identified the most suitable conditions for observation of the phase separation in Bose mixture at finite temperatures and explored the impact of intraspecies repulsion on the Bose-Einstein transition temperature for symmetric mixture. We have also calculated the temperature dependence of the Bose condensate. A controversy regarding the obtained results is that one typically thinks that the inclusion of thermal effects should destabilize the system and lead to stratification of mixture into two separate components. And this is actually the situation observed both in the mean-field and the Bogoliubov approximations. In this article, contrary, we have argued that the finite-temperature fluctuations, which are responsible for the formation of non-trivial critical behavior in the Bose-Einstein condensation point, {\it stabilize} the Bose-Bose mixture. Finally, it would be interesting to clarify how these findings correlate with experiments and results of Monte Carlo simulations.

\begin{center}
{\bf Acknowledgements}
\end{center}
Work of O.~H. was partly supported by Project FF-83F (No.~0119U002203) from the Ministry of Education and Science of Ukraine.


\begin{thebibliography}{99}
\bibitem{Petrov} D.~S. Petrov, Phys. Rev. Lett. {\bf 115}, 155302 (2015).

\bibitem{Cabrera} C.~R. Cabrera, L. Tanzi, J. Sanz, B. Naylor, P. Thomas, P. Cheiney, L. Tarruell, Science {\bf 359}, 301 (2018).
\bibitem{Semeghini} G. Semeghini, G. Ferioli, L. Masi, C. Mazzinghi, L. Wolswijk, F. Minardi, M. Modugno, G. Modugno, M. Inguscio, and M. Fattori
Phys. Rev. Lett. {\bf 120}, 235301 (2018).

\bibitem{Petrov_Astrakharchik} D.~S. Petrov and G.~E. Astrakharchik, Phys. Rev. Lett. {\bf 117}, 100401 (2016).
\bibitem{Astrakharchik_Malomed} G.~E. Astrakharchik and B.~A. Malomed,
Phys. Rev. A {\bf 98}, 013631 (2018).
\bibitem{Zin} P. Zin, M. Pylak, T. Wasak, M. Gajda, and Z. Idziaszek
Phys. Rev. A {\bf 98}, 051603(R) (2018).

\bibitem{Gautam} S. Gautam and S.~K Adhikari J. Phys. B: At. Mol. Opt. Phys. {\bf 52} 055302 (2019).

\bibitem{Cheiney} P. Cheiney, C.~R. Cabrera, J. Sanz, B. Naylor, L. Tanzi, and L. Tarruell, Phys. Rev. Lett. {\bf 120}, 135301 (2018).


\bibitem{Jorgensen} N.~B. J\o{}rgensen, G.~M. Bruun, and J.~J. Arlt
Phys. Rev. Lett. {\bf 121}, 173403 (2018).

\bibitem{Kartashov} Y.~V. Kartashov, B.~A. Malomed, L. Tarruell, and L. Torner,
Phys. Rev. A {\bf 98}, 013612 (2018).
\bibitem{Li} Y. Li, Z. Chen, Z. Luo, C. Huang, H. Tan, W. Pang, and B.~A. Malomed,
Phys. Rev. A {\bf 98}, 063602 (2018).


\bibitem{Staudinger} C. Staudinger, F. Mazzanti, and R.~E. Zillich, Phys. Rev. A {\bf 98}, 023633 (2018).
\bibitem{Cikojevic_18} V. Cikojevi\ifmmode \acute{c}\else \'{c}\fi{},  K. D\ifmmode \check{z}\else \v{z}\fi{}elalija, P. Stipanovi\ifmmode \acute{c}\else \'{c}\fi{}, L. Vranje\ifmmode \check{s}\else \v{s}\fi{} Marki\ifmmode \acute{c}\else \'{c}\fi{},  J. Boronat,  Phys. Rev. B {\bf 97}, 140502(R) (2018).
\bibitem{Cikojevic_19} V. Cikojevi\ifmmode \acute{c}\else \'{c}\fi{}, L. Vranje\ifmmode \check{s}\else \v{s}\fi{} Marki\ifmmode \acute{c}\else \'{c}\fi{}, G.~E. Astrakharchik, and J. Boronat, Phys. Rev. A {\bf 99}, 023618 (2019).
\bibitem{Utesov} O.~I. Utesov, M.~I. Baglay, and S.~V. Andreev, Phys. Rev. A {\bf 97}, 053617 (2018).
\bibitem{Konietin} P. Konietin and V. Pastukhov, J.~Low Temp.~Phys. {\bf 190}, 256 (2018).
\bibitem{Karle} V. Karle, N. Defenu, and T. Enss, Phys. Rev. A {\bf 99}, 063627 (2019).
\bibitem{Parisi} L. Parisi, G.~E. Astrakharchik, and S. Giorgini, Phys. Rev. Lett. {\bf 122}, 105302 (2019).



\bibitem{Larsen} D.~M. Larsen, Ann. Phys. {\bf 24}, 89 (1963).
\bibitem{Timmermans} E. Timmermans, Phys. Rev. Lett. {\bf 81}, 5718 (1998).

\bibitem{Balabanyan} G.~O. Balabanyan, Theor. Math. Phys. {\bf 66}, 81 (1986).
\bibitem{Oles} B.~Ole\'s and K.~Sacha, J. Phys. A: Math. Theor. {\bf 41}, 145005  (2008).
\bibitem{Vakarchuk} I.~O. Vakarchuk, V.~S. Pastukhov, J. Phys. Stud. {\bf 12}, 1001 (2008); {\it ibid} {\bf 12}, 3002 (2008).
\bibitem{Rovenchak} A. Rovenchak, Low Temp. Phys. {\bf 42}, 36 (2016).

\bibitem{Zhang} C.-H. Zhang and H.~A. Fertig, Phys. Rev. A {75}, 013601 (2007).
\bibitem{Armaitis} J. Armaitis, H.~T.~C. Stoof, and R.~A. Duine, Phys. Rev. A {\bf 91}, 043641 (2015).

\bibitem{Shi} H. Shi, W.-M. Zheng, and S.-T. Chui, Phys. Rev. A {\bf 61}, 063613 (2000).
\bibitem{Van_Schaeybroeck} B. Van~Schaeybroeck, Physica A {\bf 392}, 3806 (2013).
\bibitem{Ota} M. Ota, S. Giorgini, S. Stringari, Phys. Rev. Lett. {\bf 123}, 075301 (2019).

\bibitem{Colson} W.~B. Colson,  A.~L. Fetter, J. Low Temp. Phys. {\bf 33}, 231 (1978). 
\bibitem{Boudjemaa_18} A.~Boudjem\^aa, Phys. Rev. A {\bf 97}, 033627 (2018).

\bibitem{Hryhorchak_19(2)} O. Hryhorchak, V. Pastukhov, Physica B {\bf 583}, 412017 (2020).
\bibitem{Hryhorchak} O. Hryhorchak and V. Pastukhov, EPL (Europhysics Letters) {\bf 118}, 56003 (2017).

\bibitem{Chien} C.-C. Chien, F. Cooper, and E. Timmermans, Phys.~Rev.~A {\bf 86}, 023634 (2012).

\bibitem{Hryhorchak_19} O. Hryhorchak and V. Pastukhov, J. Phys. A: Math. Theor. {\bf 52}, 025002 (2019).
\bibitem{Nepomnyashchii} Y.~A. Nepomnyashchii, Zh. Eksp. Teor. Fiz. {\bf 70}, 1070 (1976) [Sov. Phys.—JETP {\bf 43}, 559 (1976)]; Theor. Math. Phys. {\bf 20}, 904 (1974).

\bibitem{Baym} G. Baym, J.-P. Blaizot and J. Zinn-Justin, Europhys. Lett. {\bf 49}, 150 (2000).

\bibitem{Pilati} S.~Pilati, S.~Giorgini, and N.~Prokof'ev,
Phys.~Rev.~Lett. {\bf 100}, 140405 (2008).



\end{thebibliography}
\end{document}